\documentclass[
 reprint,
amsmath,amssymb,
aps,
]{revtex4-2}

\usepackage{float}
\usepackage{longtable}
\usepackage{booktabs}
\usepackage{graphicx}
\usepackage{dcolumn}
\usepackage{bm}
\usepackage{hyperref}
\usepackage{url}
\usepackage{xcolor}
\definecolor{revisionblue}{RGB}{0,70,180}

\usepackage[linesnumbered,ruled,vlined]{algorithm2e}
\usepackage{algpseudocode}

\usepackage{url}

\begin{document}

\preprint{APS/123-QED}

\title{Bursty Arrivals, Smooth Sojourns: \\ Non-Poissonian Temporal Dynamics in a Logistics Warehouse}

\author{Alfonso de Miguel-Arribas}
\email{ademiguel@zlc.edu.es}
\affiliation{
 Zaragoza Logistics Center (ZLC), Av. de Ranillas 5, 50018, Zaragoza, Spain
}

\author{Yamir Moreno}
\affiliation{Institute for Biocomputation and Physics of Complex Systems (BIFI), University of Zaragoza, Zaragoza, 50018, Spain}
\affiliation{Department of Theoretical Physics, University of Zaragoza, Zaragoza, 50018, Spain}

\date{\today}

\begin{abstract}
Warehouses are central nodes in logistics networks: they buffer material flows, synchronize heterogeneous actors, and absorb temporal mismatches between inbound and outbound operations. Yet most warehouse analyses still rely on aggregate performance indicators or on queueing assumptions in which event timing is stationary and approximately memoryless. Here we use one month of high-resolution pallet-level data from a large Spanish warehouse to characterize arrivals, departures, and outbound residence times from a statistical-physics perspective. Inter-arrival and inter-departure times are strongly heterogeneous and compatible with heavy-tailed, non-Poissonian behavior, whereas outbound sojourn times are more naturally described by a log-normal distribution, suggesting constrained service mechanisms with a characteristic operational scale. Disaggregation by logistics flow reveals systematic differences in burstiness, memory, and distributional similarity. A renewal-based aging analysis uncovers recurrent weekly accumulation and clearance cycles in the outbound buffer zone. Finally, a Little's-Law-inspired activity--sojourn scaling identifies two operational regimes: a near-linear baseline under regular turnover and a reproducible off-baseline branch associated with weekend accumulation and Monday dispatches. These results provide a compact diagnostic framework for temporal complexity in warehouse operations and show how limited but high-resolution industrial data can reveal operational structure invisible to aggregate throughput statistics.
\end{abstract}


\maketitle


\section{Introduction}
\label{sec:intro}

In 2005, Albert-L{\'a}szl{\'o} Barab{\'a}si published a seminal article in Nature on human dynamics, presenting evidence that inter-event time distributions in human activity deviate significantly from the classical Poisson assumption ~\cite{barabasi2005origin}. He proposed a mechanism to explain the emergence of burstiness and heavy tails in daily behavioral patterns. This work sparked a surge of research---some extending the original models ~\cite{vazquez2005exact, vazquez2006modeling, goh2008burstiness, min2013burstiness}, others scrutinizing or refining them, and many applying these concepts to a variety of domains ~\cite{vazquez2007bimpact, stehle2010dynamical, moinet2015burstiness, lambiotte2013burstiness}.

Beyond the characterization of temporal structure, the tools and insights of statistical physics have been widely applied in recent decades to study human-centered and socio-technical systems. These include the Internet and the World Wide Web ~\cite{barabasi2001physics, albert2002statistical, krapivsky2002statistical, gonccalves2008human}, social dynamics ~\cite{castellano2009statistical, stehle2010dynamical}, human mobility patterns ~\cite{gonzalez2008understanding, riccardo2012towards, barbosa2018human}, transportation phenomena and systems~\cite{helbing2001traffic, schadschneider2002traffic, barthelemy2011spatial}, and urban settlements~\cite{barthelemy2019statistical, rybski2019urban}, among others. Economic systems and related phenomena have also become fertile ground for these approaches, particularly through the emergence of econophysics ~\cite{mantegna1999introduction, aoki2011reconstructing} or complexity economics ~\cite{arthur2021foundations}. Steadily, these methods have begun to permeate production, logistics, and supply chain studies ~\cite{helbing2004physics, nagatani2004stability, helbing2006understanding, ezaki2015dynamics, reisch2022monitoring, chakraborty2024inequality,   stangl2024firm}, traditionally dominated by optimization-based frameworks. However, compared to other domains, the application of complexity-oriented tools in logistics has progressed more slowly, with relatively few interdisciplinary contributions to date.

Logistics---a core function of supply chain management---concerns the coordinated flow of goods, services, and information from origin to destination. Within this domain, warehouses play a pivotal role as intermediate nodes where goods are consolidated, stored, and redistributed. Warehouse operations have traditionally been studied from the perspectives of operations research, industrial engineering, and management science ~\cite{cormier1992review, gu2007research, schwarz2015general, van2017analysis, papadopoulos2019classification, funke2020complex, kiani2020cross, boysen2023review}. These fields focus on optimizing performance metrics---such as order picking efficiency \cite{le2005design, de2007design}, space utilization ~\cite{kovacs2011optimizing, pan2018optimization, reyes2019storage}, and throughput ~\cite{cormier1992review, liu2025new}---under resource and time constraints, often relying on deterministic or stochastic optimization models ~\cite{gong2009stochastic, gong2011review, claeys2016stochastic, perez2021mathematical, teck2024simulation, boysen202550}.

While such approaches have yielded substantial insights, they typically treat event timings as memoryless or stationary, often assuming Poisson processes in queueing-theoretical frameworks ~\cite{le2007travel,roy2013blocking,gross2011fundamentals,legato2021queueing,liu2021dynamic,liu2021stochastic,lorenz2024picking}. Yet, empirical studies in human and socio-technical systems have consistently shown that inter-event times are frequently heavy-tailed, bursty (with alternating periods of high and low activity), and temporally correlated ~\cite{barabasi2005origin, vazquez2006modeling}. These features challenge the assumptions of standard queueing models and suggest the presence of hidden bottlenecks, feedbacks, or constraints invisible to aggregate metrics.

Despite their relevance, these temporal properties remain underexplored in warehouse contexts. A few pioneering studies have begun to reveal deviations from Poissonian dynamics in logistics systems. For example, Wang and Guo ~\cite{wang2010human} analyzed inbound logistics operations and found that constrained human--computer interactions produced unimodal inter-event time distributions with power-law tails. Yao et al. ~\cite{yao2014dynamic} examined outbound processes across multiple warehouses and observed bursty behavior and power-law exponents around $\gamma\approx 2.5$. Their work also revealed long-range memory (Hurst exponents $>0.5$) and fractal-like structures, further explored through multi-fractal and coupling analyses in subsequent studies ~\cite{yao2015multifractal, yao2017coupling}.

In this paper, we contribute to this scarce literature by analyzing warehouse operations as a temporal point process observed at pallet resolution. The empirical setting is necessarily specific---one facility, one operator, and one month of activity---but it is also unusually detailed for this type of industrial system. We therefore frame the paper not as a universal characterization of all warehouses, but as a proof of concept for a physics-inspired diagnostic approach that extracts interpretable temporal signatures from operational data. We focus on three linked questions. First, how far do arrival and departure processes depart from Poissonian timing? Second, do different logistics flows exhibit distinguishable temporal signatures? Third, can the time-resolved age of cargo and the activity experienced during each pallet's residence reveal operational regimes that are hidden in aggregate averages?

Our contributions are threefold. (i) We extend the small body of work connecting statistical physics and warehouse operations by analyzing real event-level data at the level at which operational decisions are recorded. (ii) We combine distributional fitting, burstiness--memory coefficients, and Jensen--Shannon comparisons to characterize global and flow-specific deviations from memoryless timing. (iii) We introduce two complementary diagnostics for the outbound buffer zone: a renewal-theoretic aging analysis that makes weekly accumulation cycles explicit, and a Little's-Law-inspired activity--sojourn scaling that separates regular turnover from calendar-driven off-baseline behavior. Together, these analyses provide a reproducible framework for identifying temporal structure, flow heterogeneity, and latent operational rhythms in warehouse data.
The rest of this paper is organized as follows. Section~\ref{sec:methods} describes the warehouse dataset, operational context, and analytical framework used to characterize the data. Section~\ref{sec:results} presents the empirical results. Section~\ref{sec:discussion} discusses implications and outlines avenues for future research.

\section{Material and methods}
\label{sec:methods}

\subsection{Data}
\label{subsec:data}

The high-resolution data used in this study were provided by a major logistics operator in Spain and record pallet circulation in a large warehouse during May 2023. Two complementary event tables were available. The first table describes inbound operations, including the arrival of pallets to the inbound buffer zone and their subsequent shelving after an initial residence period. The second table describes outbound operations, including the arrival of pallets to the outbound buffer zone, their residence in that zone, and their final departure from the warehouse. Each record is associated with a pallet identifier within its own subsystem. Because identifiers are reset and cargo may be reorganized after shelving, inbound and outbound records cannot be linked reliably at the individual-pallet level; we therefore analyze the inbound and outbound subsystems separately.

Both tables contain the timestamps required to reconstruct event sequences: arrival time at the corresponding buffer area, denoted generically by $t_{\mathrm{arr}}$, and the relevant completion time, either shelving time $t_{\mathrm{shelf}}$ or departure time $t_{\mathrm{dep}}$. The outbound table also includes the logistics flow associated with each pallet, indicating its operational route through the facility (for example, customer pickup, direct shipment, or transfer to another logistics center). The inbound table contains pallets arriving for storage and shelving and does not include the same flow categorization.

The original raw data cannot be made public because of privacy and commercial restrictions. However, the analysis code and curated derived datasets used to reproduce the figures are provided in the repository cited in the Code availability statement.
\subsection{Elementary processes at the warehouse}
\label{subsec:processes}

Throughout this work, we refer to ``processes'' as generic event types (e.g., arrivals, departures, sojourns) and ``flows'' as logistic routes defined by the cargo's operational origin and destination within the warehouse. The warehouse dynamics captured in the datasets can be described in terms of the following processes:
\begin{enumerate}
\item Inbound arrival. Pallets arrive at the warehouse either from other facilities or directly from clients and are first placed in the inbound buffer zone (IBZ). For each pallet, this process is characterized by the warehouse arrival time $t_{\mathrm{arr}}$.
\item Shelving. After a certain residence in the IBZ, pallets are transferred to shelves in the warehouse. The shelving time is denoted $t_{\mathrm{shelf}}$, and the inbound sojourn time is $\tau_{\mathrm{soj},I}=t_{\mathrm{shelf}}-t_{\mathrm{arr}}$. Not all inbound pallets are shelved; some may be routed directly to the outbound buffer zone.
\item Outbound arrival. Pallets entering the outbound buffer zone (OBZ) are registered with an outbound arrival time $t_{\mathrm{arr},O}$. At this stage, identifiers are reset relative to the inbound dataset due to cargo reorganization, preventing pallet-level linkage across the shelving transition.
\item Departure. Pallets eventually leave the warehouse from the OBZ, either toward other warehouses or directly to clients. The departure time is $t_{\mathrm{dep}}$, and the outbound sojourn time is defined as $\tau_{\mathrm{soj},O}=t_{\mathrm{dep}}-t_{\mathrm{arr},O}$.
\end{enumerate}

To further characterize the warehouse dynamics, we analyze the following inter-event quantities:
\begin{itemize}
\item Inter-arrival times, $\tau_{\mathrm{arr}}=t_{\mathrm{arr},i+1}-t_{\mathrm{arr},i}$, with $t_{\mathrm{arr},i}$ the IBZ arrival time of pallet $i$.
\item Inter-shelving times, $\tau_{\mathrm{shelf}}=t_{\mathrm{shelf},i+1}-t_{\mathrm{shelf},i}$, with $t_{\mathrm{shelf},i}$ the shelving time of pallet $i$.
\item Inter-departure times, $\tau_{\mathrm{dep}}=t_{\mathrm{dep},i+1}-t_{\mathrm{dep},i}$, with $t_{\mathrm{dep},i}$ the OBZ departure time of pallet $i$.
\end{itemize}

\subsection{Logistic flows}
\label{subsec:flows}

According to the outbound dataset and information provided by warehouse personnel, operations in this facility are organized around seven distinct logistic flows, each corresponding to a recurrent operational pathway defined by its origin, destination, and handling requirements. These are grouped into inbound and outbound flows, with each class exhibiting different temporal and operational characteristics. For clarity, we list below the original categories used by the company (translated from Spanish) together with their stylized names as adopted in this work (Figures and text). \\

\paragraph*{Inbound Flows:}
\begin{itemize}
\item Factory pickup (\texttt{FactoryPickup}): Direct collections from production plants, especially smaller factories or clients lacking their own storage infrastructure. These shipments are typically low-volume but frequent, and the warehouse functions as a consolidation hub to integrate them quickly into the broader network.
\item Inbound drag (\texttt{InboundDrag}): Shipments arriving from other warehouses or cross-docking facilities within the same logistics network. Their role is to redistribute inventory, consolidate orders, and compensate for operational imbalances across regions.
\item Picking preparation (\texttt{PickPrep}): Pallets directed internally for order assembly and preparation before joining outbound flows.
\item Full-load preparation (\texttt{FullLoadPrep}): Stock consolidated into complete truckloads, typically destined for large distribution centers or clients with high-volume needs.
\end{itemize}

\paragraph*{Outbound Flows:}
\begin{itemize}
\item Customer pickup (\texttt{CustomerPickup}): Orders collected by clients using their own transport, usually planned and scheduled in advance.
\item Direct shipments (\texttt{DirectShip}): Expeditions departing directly from the focal warehouse to final clients, without intermediate storage. These are often triggered when sufficient inventory for a single client is consolidated, or when urgency requires bypassing redistribution steps.
\item Outbound drag (\texttt{OutboundDrag}): Transfers from the focal warehouse to other logistics centers in the Iberian Peninsula, repositioning stock closer to demand or in preparation for demand peaks.
\end{itemize}

\subsection{Non-Poissonian dynamics: Burstiness and memory}
\label{subsec:nonpoisson}

In a homogeneous Poisson process, events occur independently at a constant rate $\lambda$, and inter-event times follow the exponential density $P_{\mathrm{P}}(\tau)=\lambda \exp(-\lambda \tau)$. This provides the standard memoryless benchmark for many queueing descriptions. Numerous natural and socio-technical systems, however, display strongly inhomogeneous activity: events concentrate in bursts of high activity separated by long periods of relative inactivity \cite{jo2015correlated}. Such burstiness is a direct signature of departures from homogeneous Poisson timing.

A common empirical signature of non-Poissonian dynamics is a broad inter-event-time distribution, often approximated over part of its support by
\begin{equation}
P(\tau) \sim \tau^{-\gamma},
\end{equation}
where $\tau$ is the time between consecutive events. We use this expression as a scaling model for the tail of the distribution, not as an a priori claim that all observed variability is generated by a pure power law over the full domain. The distributional shape alone, moreover, does not determine the temporal organization of the sequence. Following Goh and Barab{\'a}si \cite{goh2008burstiness}, we therefore complement tail fits with two summary statistics: memory and burstiness.

The memory coefficient $M$ quantifies correlations between consecutive inter-event times,
\begin{equation}
M = \frac{\langle (\tau_i - \langle \tau \rangle)(\tau_{i+1} - \langle \tau \rangle) \rangle}{\sigma^2},
\label{eq:memory}
\end{equation}
where $\langle \tau \rangle$ and $\sigma^2$ are the mean and variance of the inter-event times. Positive values indicate persistence, with short intervals tending to follow short intervals and long intervals tending to follow long intervals; negative values indicate alternation.

The burstiness coefficient $B$ is the normalized dispersion
\begin{equation}
B = \frac{\sigma-\langle \tau \rangle}{\sigma+\langle \tau \rangle}.
\label{eq:burstiness}
\end{equation}
It ranges from $B=-1$ for perfectly regular timing to $B=1$ for extremely bursty timing, while $B=0$ corresponds to the exponential benchmark. In this study (Sec.~\ref{subsec:interevents}), we compute $B$ and $M$ for arrival, departure, and sojourn sequences, both globally and after disaggregation by logistics flow. These indicators are intentionally low-dimensional: they do not replace the full distributional analysis, but they make it possible to compare many operational sequences in a common phase diagram.
\subsection{Renewal theory and cargo aging}
\label{subsec:renewal}

To characterize temporal imbalances inside the warehouse, we analyze the age structure of cargo present in the outbound buffer zone over calendar time. This view is inspired by renewal theory: each pallet enters the buffer, spends a residence time in it, and leaves when the corresponding outbound operation is completed.

Let $t_{\mathrm{arr},i}$ and $t_{\mathrm{dep},i}$ denote the arrival and departure times of pallet $i$ in the outbound subsystem. Its age at calendar time $t$ is
\begin{equation}
    \tau_{\mathrm{age},i}(t) = t - t_{\mathrm{arr},i}, \quad \text{for } t_{\mathrm{arr},i} \leq t < t_{\mathrm{dep},i}.
\end{equation}
Thus $\tau_{\mathrm{age},i}(t)$ is the elapsed time since the pallet entered the buffer, conditional on the pallet still being present. Once $t\geq t_{\mathrm{dep},i}$, the completed residence time is the sojourn time $\tau_{\mathrm{soj},i}=t_{\mathrm{dep},i}-t_{\mathrm{arr},i}$.

At any time $t$, the active set of cargo is
\begin{equation}
A(t)=\{i: t_{\mathrm{arr},i}\leq t < t_{\mathrm{dep},i}\},
\end{equation}
and $N(t)=|A(t)|$ is the instantaneous number of pallets in the buffer. The empirical age distribution is then
\begin{equation}
    P(\tau_{\mathrm{age}}; t) = \frac{1}{N(t)} \sum_{i \in A(t)} \delta\!\left(\tau_{\mathrm{age}} - (t - t_{\mathrm{arr},i})\right).
    \label{eq:age_distribution}
\end{equation}
This distribution tells us whether the current buffer population is dominated by recently arrived pallets or by cargo that has been waiting for a long time. In a stationary renewal process with constant input rate and independent, identically distributed sojourns, the age distribution would converge to a time-independent equilibrium form. In the warehouse, by contrast, scheduled workflows and weekend interruptions make $P(\tau_{\mathrm{age}};t)$ explicitly time dependent. This time-resolved aging perspective complements the inter-event analysis by describing the instantaneous operational state of the buffer rather than only the completed event sequence.

\section{Results}
\label{sec:results}

In this section, we present the empirical findings derived from one month of pallet-level event data. The analysis proceeds through four complementary perspectives. We first characterize the distributions of inter-event times for the main warehouse processes, globally and by logistics flow. We then summarize temporal organization using memory and burstiness coefficients. Next, we examine the age and turnover of cargo in the outbound buffer zone using the renewal framework introduced above. Finally, we relate the sojourn time of each pallet to the amount of activity observed during its stay, using Little's Law as a macroscopic reference rather than as a literal microscopic identity.

\subsection{Inter-event time statistics}
\label{subsec:interevents}

\subsubsection{Global processes}

We begin our empirical analysis by characterizing the timing of key warehouse events. Specifically, we study the inter-event time distributions for (i) inbound buffer zone (IBZ) cargo arrivals, (ii) shelf-racking transitions, (iii) outbound buffer zone (OBZ) arrivals, (iv) OBZ departures, and (v) OBZ sojourn durations. These represent the fundamental temporal building blocks governing cargo transitions within the warehouse system.

\begin{figure}
\centering
    \includegraphics[width=\columnwidth]{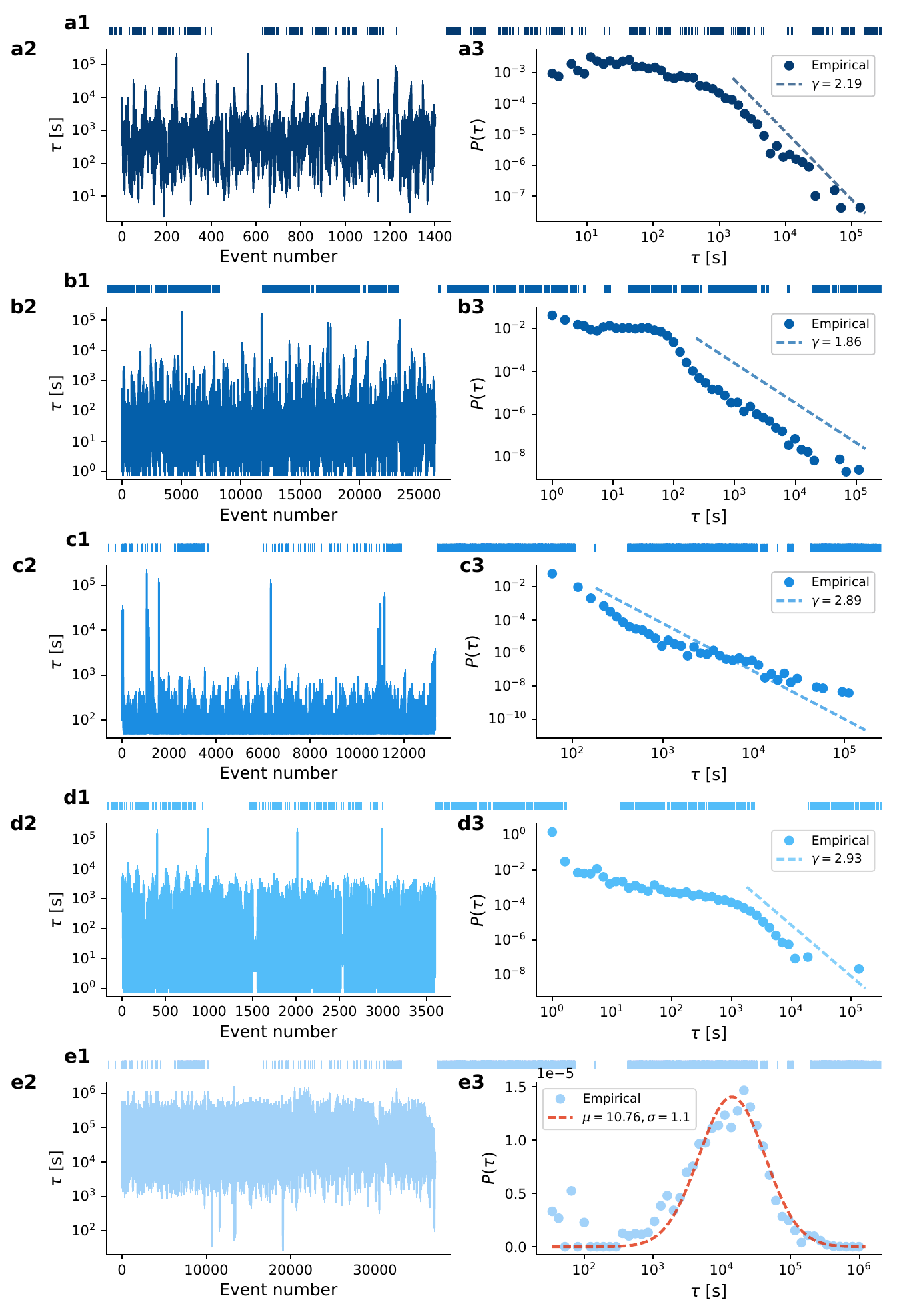}
    \caption{
    \textbf{Non-Poissonian behavior of warehouse dynamics}. The rows describe, respectively, (\textbf{a}) IBZ cargo inter-arrivals, (\textbf{b}) inter-shelf-racking times, (\textbf{c}) OBZ cargo inter-arrivals, (\textbf{d}) OBZ cargo inter-departures, and (\textbf{e}) OBZ cargo sojourn times. Panels (\textbf{a1}--\textbf{e1}) show barcode plots of the event sequences; panels (\textbf{a2}--\textbf{e2}) show delays between chronologically ordered events; panels (\textbf{a3}--\textbf{e3}) show the corresponding empirical distributions $P(\tau)$. Power-law tail fits are shown for the inter-event distributions, while the sojourn-time distribution is compared with a log-normal fit. The fitted values shown in the panels are: inbound inter-arrivals, $\gamma=2.19$; shelf-racking intervals, $\gamma=1.86$; outbound inter-arrivals, $\gamma=2.89$; outbound inter-departures, $\gamma=2.93$; and outbound sojourns, $\mu=10.76$, $\sigma=1.10$. All processes span one month of activity.}
    \label{fig:nonpoisson}
\end{figure}

Figure~\ref{fig:nonpoisson} presents a multi-panel overview of these processes. Each row corresponds to a different event type. Within each row, the top panel (a1--e1) shows barcode plots representing the raw sequence of events over time. The left column (a2--e2) plots delays between chronologically consecutive events, revealing heterogeneous rhythms and bursty dynamics. The right column (a3--e3) shows the corresponding inter-event time distributions $P(\tau)$ on log-log axes.

All distributions span several orders of magnitude, from seconds to days. The inter-event distributions decrease monotonically and display broad tails, whereas the sojourn-time distribution is unimodal and is better summarized by a log-normal form. We therefore interpret arrivals and departures as bursty point processes and sojourns as residence times generated by a more constrained service mechanism.

Using the \texttt{powerlaw} Python package \cite{alstott2014powerlaw}, we estimate scaling exponents by maximum likelihood. The lower cutoff $\tau_0$ is selected by minimizing the Kolmogorov--Smirnov distance between the empirical and fitted tail distributions \cite{clauset2009power}. The fitted exponents shown in Fig.~\ref{fig:nonpoisson} are $\gamma_{\mathrm{I,arr}}=2.19$, $\gamma_{\mathrm{shelf}}=1.86$, $\gamma_{\mathrm{O,arr}}=2.89$, and $\gamma_{\mathrm{O,dep}}=2.93$. Because the observation window is short and warehouse activity is strongly calendar-dependent, these estimates should be read as evidence for heavy-tailed, non-Poissonian timing rather than as definitive proof of asymptotic power-law behavior. Outbound sojourns, in contrast, exhibit a clear mode and are well represented by a log-normal distribution with $\mu=10.76$ and $\sigma=1.10$.

Operationally, most pallets remain in the OBZ between $10^3$ and $10^5$ seconds, i.e., from tens of minutes to approximately one day, with a pronounced concentration around the 24-hour scale. Approximately $70\%$ of completed outbound sojourns end within 24 hours, whereas $28.7\%$ exceed that threshold.

\paragraph*{Mechanistic interpretation.}
The emergence of heavy-tailed inter-event distributions has been widely documented across complex systems \cite{karsai2018bursty}. Candidate mechanisms include prioritization, hierarchical organization, heterogeneous constraints, and external modulation \cite{newman2005power}. In human dynamics, in particular, cascading Poisson processes with circadian or weekly activation windows can generate bursty activity and apparently scale-free inter-event times \cite{malmgren2008poissonian, blenn2016human}. This point is relevant here because the warehouse is not continuously operated at a constant rate: daily schedules and weekend interruptions can broaden inter-event distributions even when the underlying within-shift dynamics are much less heterogeneous.

The sojourn-time distribution points to a different mechanism. Its log-normal shape is consistent with empirical evidence that human-mediated service and response times often follow log-normal statistics \cite{sheridan2013human, blenn2016human, gualandi2018call, gualandi2019human}. A standard interpretation is multiplicative: the realized residence time accumulates many small proportional delays, such as handling variation, release constraints, order consolidation, dock availability, and transport scheduling \cite{mitzenmacher2004brief}. A more behavioral explanation is provided by Gualandi and Toscani \cite{gualandi2019human}, who show that bounded adjustment of effort around a soft service target can lead to a Fokker--Planck description whose stationary solution is log-normal. This interpretation fits the operational setting: pallets are subject to expected delivery windows and service priorities, which constrain the upper tail of residence times while still allowing multiplicative micro-delays.
\begin{figure*}
\centering
    \includegraphics[width=1.0\textwidth]{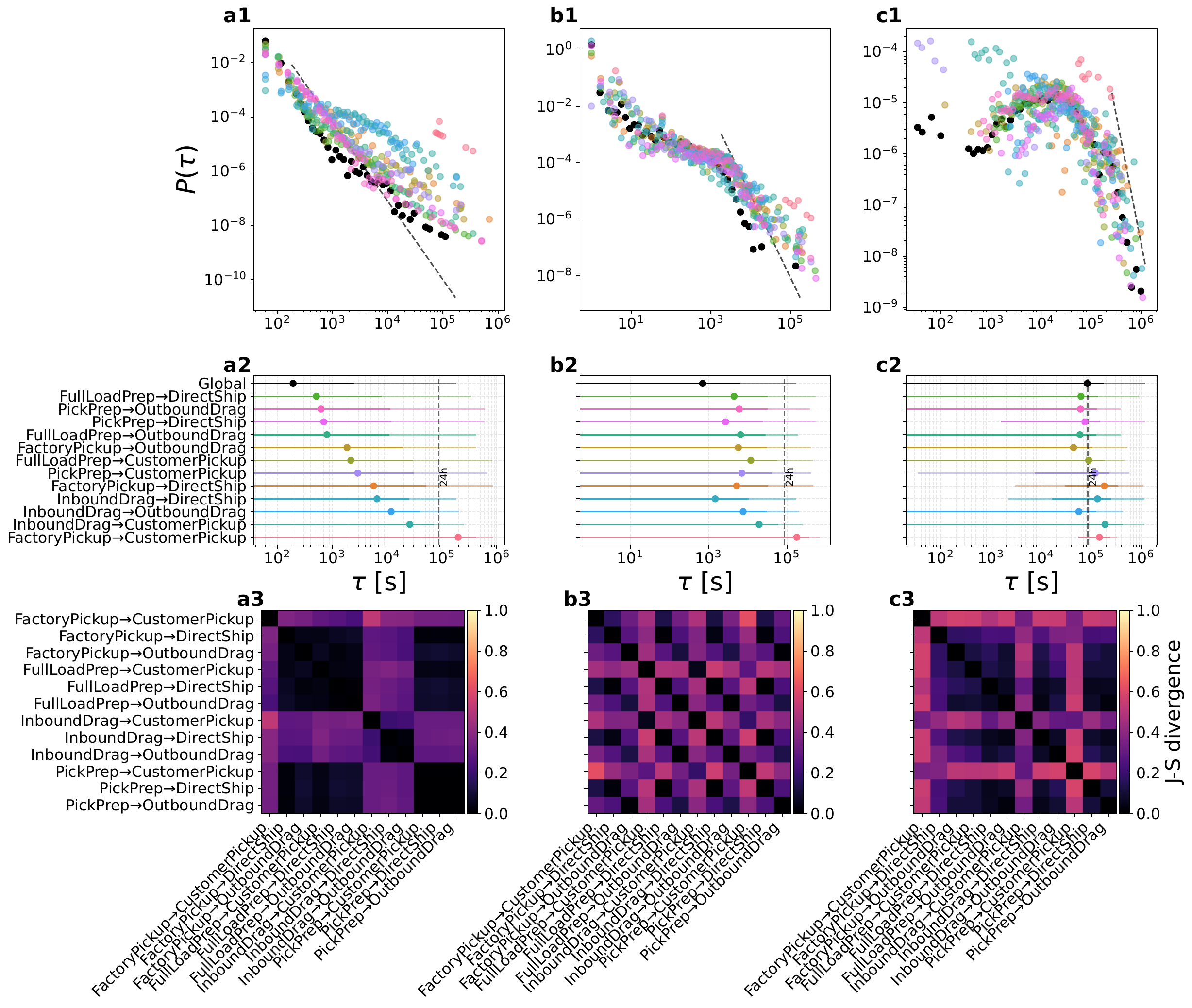}
    \caption{
    \textbf{Inter-event distributions disaggregated by logistics flow}. Top panels show empirical distributions for (\textbf{a1}) outbound inter-arrival times, (\textbf{b1}) outbound inter-departure times, and (\textbf{c1}) outbound sojourn times, both globally (black dots) and by logistics flow. Middle panels show summary statistics for the same quantities: mean (dot), one standard deviation (thick line), and min--max range (thin line). The vertical dashed line marks 24 hours. Flows are sorted by increasing mean inter-arrival time. Bottom panels (\textbf{a3}, \textbf{b3}, \textbf{c3}) show pairwise Jensen--Shannon divergence between flow-specific distributions; here flows are sorted alphabetically.}
    \label{fig:flows}
\end{figure*}

\subsubsection{Disaggregation per logistic flows}

We next disaggregate the outbound dynamics by logistics flow. Figure~\ref{fig:flows} compares global and flow-specific distributions for OBZ inter-arrivals, inter-departures, and sojourns. The middle panels summarize mean values, standard deviations, and ranges, whereas the lower panels quantify distributional similarity using Jensen--Shannon divergence. This disaggregation is useful because the global distribution mixes routes with different volumes, priorities, and scheduling constraints.

Inter-arrival times show the largest flow-to-flow variability. High-volume preparation and direct-shipment routes, such as \texttt{FullLoadPrep}$\to$\texttt{DirectShip}, \texttt{PickPrep}$\to$\texttt{OutboundDrag}, and \texttt{PickPrep}$\to$\texttt{DirectShip}, lie below the global mean inter-arrival time, consistent with frequent activation. In contrast, \texttt{FactoryPickup}$\to$\texttt{CustomerPickup} has a mean inter-arrival time exceeding 24 hours and is therefore clearly separated from the high-volume routes. Inter-departure times are more tightly clustered across flows, suggesting stronger synchronization by dispatch schedules. Sojourn times are centered near the 24-hour scale for many flows, but their dispersion and tail behavior differ substantially, indicating that similar average residence times can coexist with very different operational risk profiles.

To quantify these differences beyond summary statistics, we compute pairwise Jensen--Shannon (J--S) divergence between flow-specific empirical distributions (Fig.~\ref{fig:flows}, bottom row). For two discrete distributions $P$ and $Q$, the J--S divergence is \cite{lin2002divergence}
\begin{equation}
\mathrm{JS}(P, Q) = \frac{1}{2} D_{\mathrm{KL}}(P \parallel R) + \frac{1}{2} D_{\mathrm{KL}}(Q \parallel R),
\label{eq:js}
\end{equation}
where $R=(P+Q)/2$, and
\begin{equation}
D_{\mathrm{KL}}(P \parallel Q) = \sum_i P(i) \log \frac{P(i)}{Q(i)}.
\label{eq:kl}
\end{equation}
Unlike $D_{\mathrm{KL}}$, the J--S divergence is symmetric and finite, making it a useful descriptive measure of temporal similarity between flows.

For inter-arrivals, most pairs show low divergence (J--S $\in[0,0.2]$), but a subset of routes is systematically more dissimilar, notably \texttt{FactoryPickup}$\to$\texttt{CustomerPickup} and the \texttt{InboundDrag}-originating routes. This suggests that \texttt{InboundDrag} arrivals are governed by a distinct replenishment or transfer logic. Inter-departures show a more heterogeneous divergence landscape, consistent with the stronger role of flow-specific dispatch windows. Sojourn distributions again separate into a broadly similar group and a more distinctive set involving \texttt{CustomerPickup}, indicating that customer-controlled collection introduces additional timing constraints not present in standard outbound transfers.

\subsection{Memory and Burstiness diagram}
\label{subsec:memory_bursty}

\begin{figure}
\centering
    \includegraphics[width=\columnwidth]{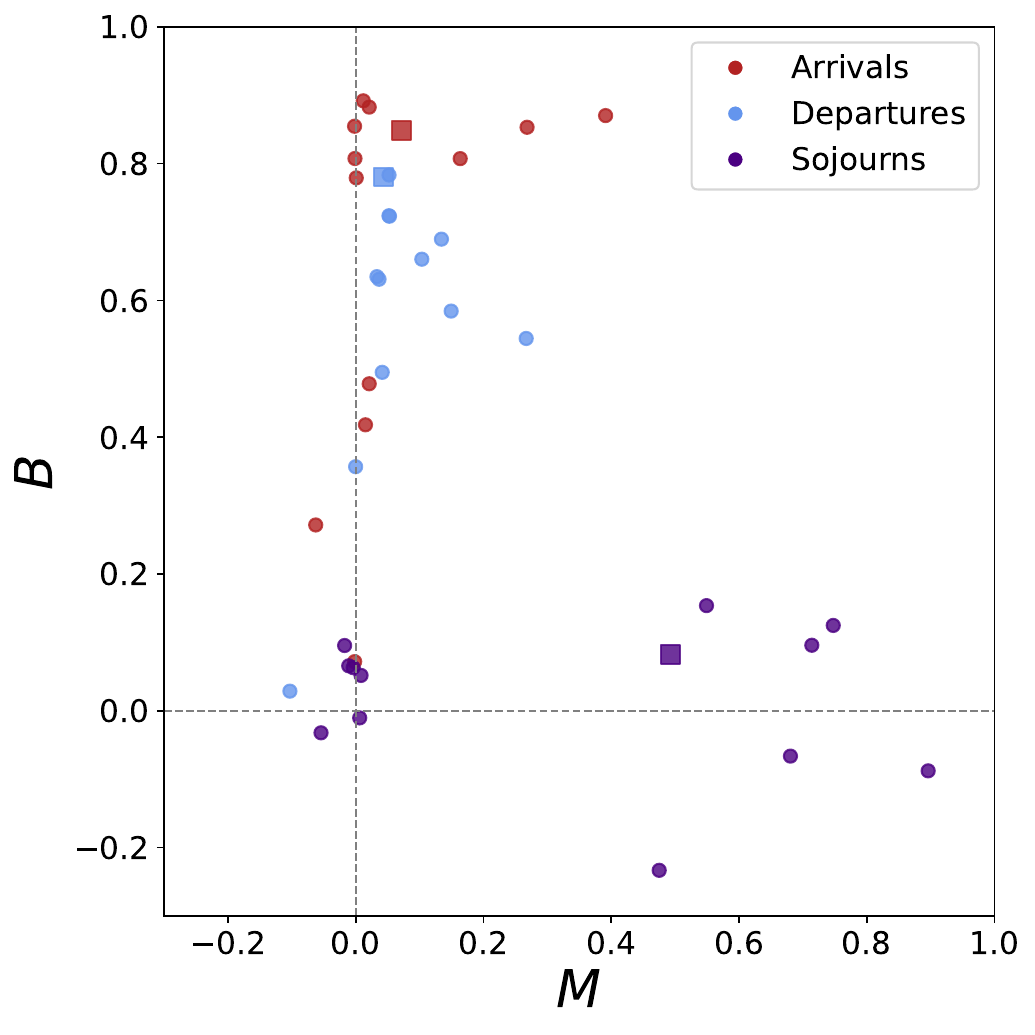}
    \caption{
    \textbf{Memory-Burstiness diagram}. Cargo-related event sequences (i.e., inter-arrivals, inter-departures, and sojourns) are characterized by their memory coefficient $M$ and burstiness coefficient $B$, following \cite{goh2008burstiness}. Global-level statistics are shown as squares, and per-flow sequences are shown as dots.
    }
    \label{fig:mb_diagram}
\end{figure}

To summarize the structure of these temporal processes more systematically, we compute the memory ($M$) and burstiness ($B$) coefficients as defined in Section~\ref{subsec:nonpoisson}. Figure~\ref{fig:mb_diagram} displays all cargo-related time series on the $M$--$B$ phase diagram. Global-level sequences are represented by squares, while flow-specific sequences appear as dots, revealing a broad range of temporal signatures across event types and operational roles.

Clear structural differences emerge between inter-event and sojourn dynamics. The global inter-arrival and inter-departure sequences occupy similar regions of the $M$--$B$ plane, with high burstiness ($B_{\mathrm{arr}}=0.85$, $B_{\mathrm{dep}}=0.78$) and near-zero memory ($M_{\mathrm{arr}}=0.07$, $M_{\mathrm{dep}}=0.04$). They are therefore highly irregular but only weakly correlated at the level of consecutive intervals. By contrast, sojourn times have low burstiness ($B=0.08$) and moderate positive memory ($M=0.49$), indicating smoother residence times with temporal persistence.

Flow-level points refine this picture. Most inter-arrival sequences remain highly bursty with little memory. Inter-departures are also bursty but display somewhat greater flow-dependent variation, likely reflecting dispatch schedules. Sojourn sequences separate into two groups: one near the origin, consistent with weakly structured residence times, and another with moderate-to-high memory but weak burstiness, consistent with runs of similarly delayed or similarly fast cargo. Thus, the $M$--$B$ diagram compresses the distributional heterogeneity of Fig.~\ref{fig:flows} into an interpretable map of temporal organization.

These results support the central empirical claim of the paper: warehouse timing is not well described by a single memoryless process. Arrivals and departures are bursty and broadly distributed, whereas residence times are smoother, more constrained, and in some flows temporally correlated. This contrast suggests two coupled layers of dynamics: external or schedule-driven activation of events, and internal service rules that regulate the time cargo spends in the buffer.

For a comprehensive statistical summary of all warehouse processes, including global and flow-specific metrics, we refer the reader to Table~\ref{tab:summary} in Appendix~\ref{app:summary}.

\subsection{Aging (Cargo turnover)}
\label{subsec:aging}

To complement the previous time-integrated analyses, we examine the age structure of cargo within the OBZ over calendar time. Drawing from renewal theory (see Sec.~\ref{subsec:renewal}), we define the age of a pallet at time $t$ as the time elapsed since its arrival to the OBZ, i.e., $\tau_{\text{age}} = t - t_{\text{arr}}$, for $t_{\text{arr}} \leq t < t_{\text{dep}}$.

In classical renewal systems, the age distribution $P(\tau_{\text{age}}; t)$ characterizes the time since the last renewal (in this case, cargo arrival) for all active elements at time $t$. When input rates are stationary and sojourns are i.i.d., this distribution converges to an equilibrium form related to the survivor function of the sojourn time distribution. Here, by contrast, we empirically trace $P(\tau_{\text{age}}; t)$ over real calendar time, capturing how bursty inflows, processing delays, and operational cycles shape the warehouse's internal state. In practice, we compute $P(\tau_{\text{age}}; t)$ following the empirical definition introduced in Eq.~\ref{eq:age_distribution}, aggregating the age values of all active pallets present at each time $t$. 

\begin{figure}
\centering
    \includegraphics[width=\columnwidth]{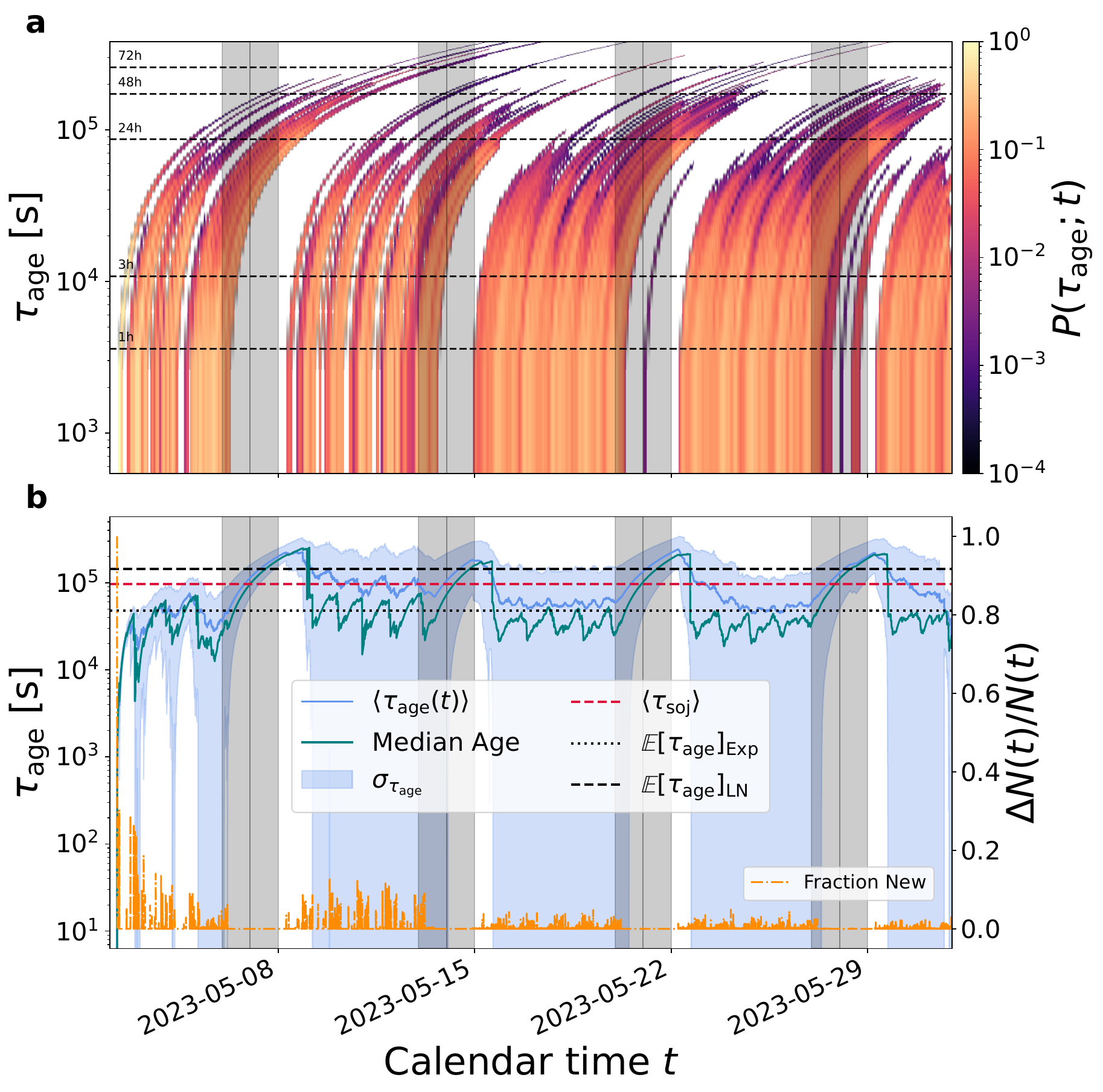}
    \caption{
    \textbf{Cargo aging and renewal}. Panel \textbf{a}: empirical distribution of cargo age $\tau_{\mathrm{age}}$ as a function of calendar time, $P(\tau_{\mathrm{age}};t)$, computed at 60-second resolution. Horizontal dashed lines mark 1 h, 3 h, 24 h, 48 h, 72 h, and 1 week. Gray vertical bands mark weekends. Panel \textbf{b}: time-resolved summary statistics of $P(\tau_{\mathrm{age}};t)$: mean age $\langle\tau_{\mathrm{age}}(t)\rangle$, median age, and standard deviation $\sigma_{\tau_{\mathrm{age}}}$. The red dashed line marks the average completed sojourn time $\langle\tau_{\mathrm{soj}}\rangle$. The black dotted line shows the exponential-renewal expectation $\mathbb{E}[\tau_{\mathrm{age}}]_{\mathrm{Exp}}=\langle\tau_{\mathrm{soj}}\rangle/2$, and the black dashed line shows the log-normal renewal expectation $\mathbb{E}[\tau_{\mathrm{age}}]_{\mathrm{LN}}$. The right axis reports the arrival-to-stock ratio $\Delta N(t)/N(t)$.}
    \label{fig:aging}
\end{figure}

Figure~\ref{fig:aging}a displays the empirical age distribution $P(\tau_{\text{age}}; t)$ over the month-long observation window, with the age axis in logarithmic scale. Weekend periods are shaded in gray. The distribution evolves dynamically over time, reflecting both operational cycles and burst-driven fluctuations.

At the beginning of each working week, the active cargo population is dominated by recently arrived pallets, with ages mostly below a few hours. As the week progresses, the distribution stretches toward larger ages: new pallets continue to enter the buffer, but a fraction of slower-moving cargo remains and accumulates residence time. During weekends, inbound and outbound activity slow down substantially, and the age distribution concentrates above the 24-hour scale.

The restart of activity on Mondays produces a recurrent bimodal pattern. One component corresponds to aged cargo that entered before the weekend and has not yet departed; the other corresponds to newly arriving pallets. As Monday and Tuesday operations clear the accumulated stock, the older component dissipates and the age distribution moves back toward younger values. This weekly aging--clearance cycle is one of the clearest operational signatures in the dataset.

Figure~\ref{fig:aging}b summarizes this evolution through the mean age, median age, standard deviation, and arrival-to-stock ratio $\Delta N(t)/N(t)$. The mean and median initially evolve together, but they separate when a small fraction of long-residence cargo pulls the mean upward while the median remains controlled by faster-turning pallets. This separation is a population-level manifestation of the waiting-time paradox: at a randomly chosen calendar time, long-lived items are overrepresented among the cargo currently present in the system.

Under stationarity and independent, identically distributed sojourns, renewal theory gives the mean age
\begin{equation}
\mathbb{E}[\tau_{\mathrm{age}}] = \frac{\mathbb{E}[\tau_{\mathrm{soj}}^2]}{2\mathbb{E}[\tau_{\mathrm{soj}}]}.
\end{equation}
For exponential sojourns, this reduces to $\mathbb{E}[\tau_{\mathrm{age}}]=\mathbb{E}[\tau_{\mathrm{soj}}]/2$. For log-normal sojourns with parameters $\mu$ and $\sigma$,
\begin{align}
\mathbb{E}[\tau_{\mathrm{soj}}] &= e^{\mu + \sigma^2/2}, \\
\mathbb{E}[\tau_{\mathrm{soj}}^2] &= e^{2\mu + 2\sigma^2}, \\
\mathbb{E}[\tau_{\mathrm{age}}] &= \frac{1}{2} e^{\mu + 3\sigma^2/2}.
\end{align}
Using the fitted log-normal parameters, these benchmarks satisfy
\begin{equation}
\mathbb{E}[\tau_{\mathrm{age}}]_{\mathrm{LN}} > \langle \tau_{\mathrm{soj}} \rangle > \mathbb{E}[\tau_{\mathrm{age}}]_{\mathrm{Exp}}=\frac{\langle \tau_{\mathrm{soj}} \rangle}{2}.
\end{equation}
The empirical mean age does not remain at either stationary benchmark. Instead, it oscillates with the weekly operating cycle, moving downward during high-turnover periods and upward during weekend accumulation. This mismatch is informative rather than problematic: it shows that the warehouse is a non-stationary renewal system whose apparent equilibrium depends on calendar time.

\begin{figure*}
\centering
    \includegraphics[width=1.0\textwidth]{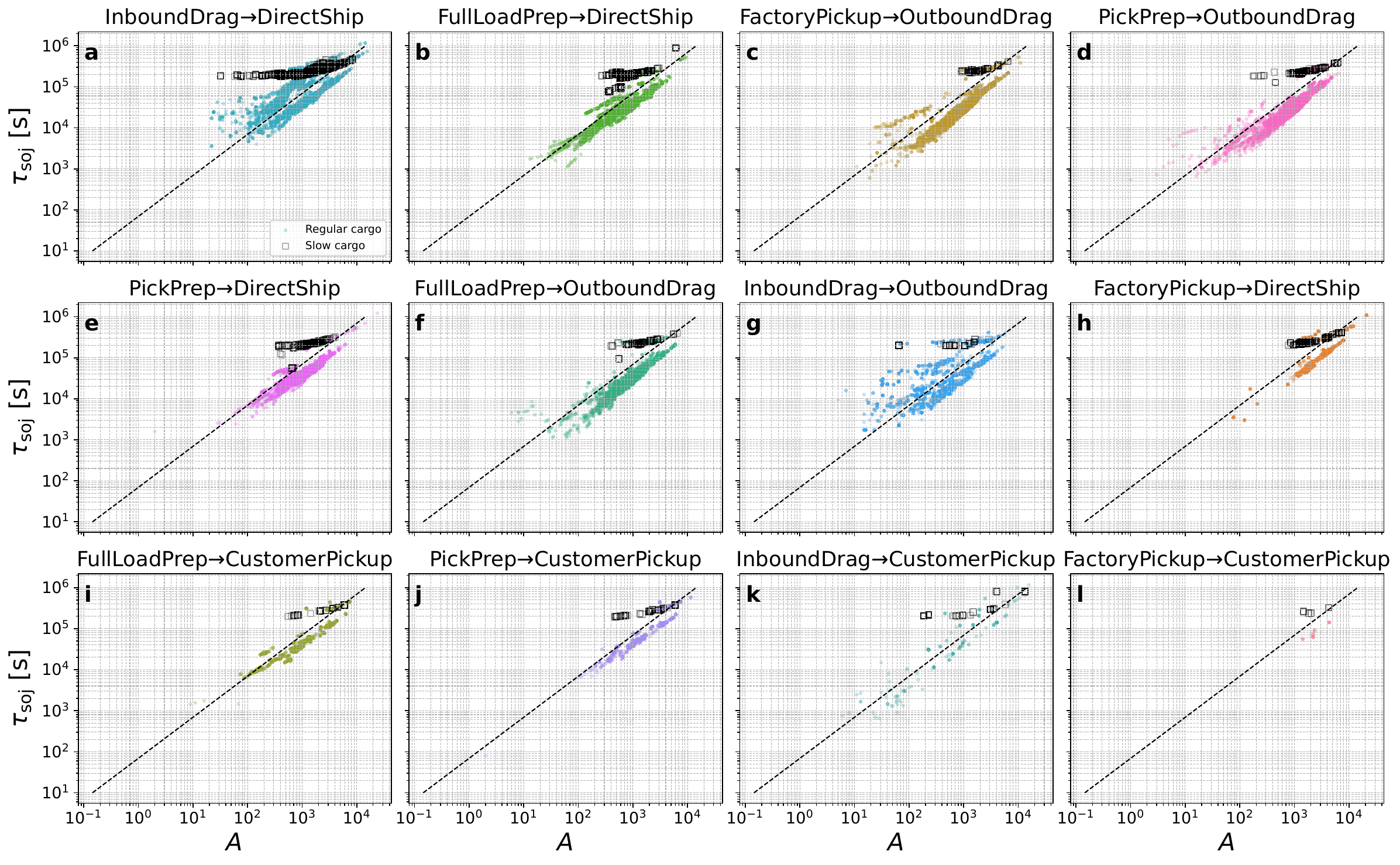}
    \caption{
    \textbf{Activity scaling.} Log--log scatter plots of outbound sojourn time $\tau_{\mathrm{soj},i}$ versus experienced activity $A_i$, defined as the number of OBZ arrivals occurring during pallet $i$'s stay. Each panel corresponds to one logistics flow, ordered by total volume. Colored points are individual pallets; black square outlines mark cargo that arrived before a weekend and departed on a Monday. The dashed black line is the null benchmark $\tau=A/\lambda$, where $\lambda$ is the average arrival rate over the observation window. Most pallets lie close to a near-linear baseline, while a reproducible off-baseline branch is associated with weekend accumulation and Monday clearance.}
    \label{fig:scaling}
\end{figure*}

Altogether, the aging analysis provides a time-resolved, population-level view of warehouse turnover. It complements the static sojourn distribution by showing when long residence times are created and cleared. The result is not merely that some pallets stay longer than others, but that long residences are organized by the weekly rhythm of the warehouse.

\subsection{Activity scaling}
\label{subsec:scaling}

\paragraph*{Theoretical background and null model.}
Little's Law relates the long-run average number of items in a stable system to the arrival rate and the average residence time \cite{little2008little}:
\begin{equation}
    \langle N\rangle=\lambda\langle\tau\rangle.
\end{equation}
This is a macroscopic identity. We do not assume that it holds pallet by pallet. Instead, we use it to define a simple null benchmark for the relationship between a pallet's residence time and the amount of traffic observed while it is present. For pallet $i$, let
\begin{equation}
    A_i=N_{\mathrm{arr}}\!\left([t_{\mathrm{arr},i},t_{\mathrm{dep},i})\right)
\end{equation}
be the number of OBZ arrivals during its sojourn. If the arrival process were approximately stationary at rate $\lambda$, then the expected activity during a residence time $\tau_i$ would scale as $A_i\simeq \lambda \tau_i$. Equivalently, a natural null estimate for the residence time associated with an observed activity $A_i$ is
\begin{equation}
    \tau_i^{\mathrm{null}}=\frac{A_i}{\lambda},\qquad \lambda=\frac{N_{\mathrm{arr}}}{T},
\end{equation}
where $T$ is the total observation window. In a log--log plot, this benchmark is a diagonal line with slope one. Deviations from it identify periods in which calendar effects, service interruptions, or flow-specific rules decouple elapsed time from the amount of concurrent activity.

\paragraph*{Empirical observations.}
Figure~\ref{fig:scaling} shows the relationship between $\tau_{\mathrm{soj},i}$ and $A_i$ for each outbound pallet, disaggregated by logistics flow. Across flows, most pallets form a near-linear band close to the null benchmark. This indicates that, during regular operation, elapsed residence time and experienced activity increase approximately proportionally. The alignment is not exact, and different flows show different dispersion and intercepts, but the main regime is consistent with quasi-stationary turnover.

A second, reproducible branch appears above the baseline in almost every sufficiently populated flow. These points have residence times in the range $10^5$--$10^6$ seconds and are overwhelmingly associated with pallets that entered before a weekend and departed on Monday. Their activity footprint $A_i$ is not exceptionally high; rather, elapsed time increases while service is partially interrupted. Thus, the off-baseline branch is not simply a congestion effect caused by many arrivals. It is a calendar-driven stagnation effect: time accumulates even when the activity clock advances slowly.

To quantify this structure, we classify pallets as off-baseline when
$\log(\tau_i/\tau_i^{\rm null})>\log 2$ and $\tau_i>24\,\mathrm{h}$,
where $\tau_i^{\rm null}=A_i/\lambda$. This residual-based definition does not use the day of departure. Appendix Fig.~\ref{fig:summary} shows that off-baseline pallets are nevertheless strongly concentrated among Monday departures in most flows, and that their mean excess delay is positive across all reported flows.

The scaling plot therefore separates two operational regimes. In the regular regime, residence time is largely explained by the volume of activity experienced during the pallet's stay. In the weekend-accumulation regime, residence time is inflated by scheduled inactivity, producing a systematic deviation from the baseline. This distinction is operationally useful because it separates delays caused by high traffic from delays caused by temporal discontinuities in service. Both mechanisms increase sojourn time, but they suggest different interventions: capacity or staffing adjustments in the former case, and schedule coordination or pre-weekend clearance rules in the latter.

\section{Discussion}
\label{sec:discussion}

This study analyzed the temporal dynamics of cargo movement in a real logistics warehouse using tools from statistical physics and renewal theory. The dataset is limited to one warehouse, one operator, and one month, but it is recorded at high temporal resolution and at the operational unit of interest: the pallet. This makes it possible to move beyond aggregate throughput and to ask how arrivals, departures, and residence times are organized in time.

The main empirical message is that the warehouse is not well represented by a homogeneous Poisson picture. Inter-arrival and inter-departure sequences are highly heterogeneous and bursty, with broad tails compatible with non-Poissonian timing. Outbound sojourns, by contrast, are smoother and are better described by a log-normal residence-time distribution, suggesting the action of bounded service mechanisms, delivery windows, and multiplicative micro-delays. Disaggregation by logistics flow shows that these temporal signatures are not uniform: routes involving \texttt{CustomerPickup} or \texttt{InboundDrag} often differ from the global pattern, indicating that operational origin and destination matter for the timing statistics.

The renewal and scaling analyses provide the clearest operational interpretation. The age distribution of active cargo reveals a recurrent weekly cycle in which pallets accumulate age during weekends and are cleared when activity resumes. The activity--sojourn scaling then shows how this calendar effect appears at the microscopic level: most pallets follow a near-linear activity baseline, but pre-weekend cargo dispatched on Mondays forms a systematic off-baseline branch. In practical terms, the analysis distinguishes delays associated with regular traffic from delays caused by scheduled interruptions. This distinction is difficult to see from average sojourn times alone.

The study also has clear limitations. The short observation window prevents strong claims about universality, seasonality across months, or asymptotic scaling exponents. Some flow-specific subsets are small, so their fitted exponents and divergence values should be interpreted cautiously. Moreover, the lack of spatial information prevents us from linking temporal signatures to storage location, travel distance, dock assignment, or picker routing. For these reasons, our claims are deliberately framed as diagnostic and mechanistic rather than universal. The value of the paper lies in showing that even a limited but high-resolution industrial dataset contains measurable temporal structure that is meaningful for both complex-systems analysis and warehouse management.

Future work should extend the analysis in three directions. First, longer and multi-site datasets are needed to test whether the signatures reported here---bursty activation, log-normal residence times, flow-specific temporal fingerprints, and weekly aging cycles---are stable across warehouses, firms, sectors, and seasons. Second, richer models should combine non-stationary arrival processes with flow-specific service rules and calendar-dependent capacity. Third, spatially annotated data would allow temporal diagnostics to be connected with layout, storage policy, and intra-warehouse movement. Such extensions could help bridge the gap between statistical physics and operations research by turning event-level data into interpretable diagnostics for complex supply-chain nodes.

\subsection*{Code availability}
\label{subsec:code}

The code developed for data analysis and figure generation is hosted in the repository cited in Ref.~\cite{warehouse_repo}.

\begin{acknowledgments}
    The authors thank the logistics operator, \textit{Carreras Grupo Log{\'i}stico}, for providing the data and for sharing operational knowledge that made the interpretation of the results possible. The authors are also grateful to Carolina Cipr\'es for her support in facilitating the institutional arrangements that enabled the study. The authors thank Alejandro Tejedor for helpful feedback on an earlier version of the manuscript. Y.M. was partially supported by the Government of Arag\'on, Spain, and ERDF ``A way of making Europe'' through grant E36-23R (FENOL), and by Grant No. PID2023-149409NB-I00 from Ministerio de Ciencia, Innovaci\'on y Universidades, Agencia Espa\~nola de Investigaci\'on (MICIU/AEI/10.13039/501100011033) and ERDF ``A way of making Europe''.
\end{acknowledgments}

\bibliography{references}

\clearpage

\appendix

\setcounter{figure}{0}
\renewcommand{\thefigure}{S\arabic{figure}}
\setcounter{table}{0}
\renewcommand{\thetable}{S\arabic{table}}

\onecolumngrid

\section{Summary statistics of warehouse processes}
\label{app:summary}

\begin{longtable}{llrrrrrrrr}

\caption[]{Summary statistics of warehouse processes by logistics flow. For each process and flow type, we report the number of data points, tail-fit parameters when applicable, the scaling onset $\tau_0$, mean inter-event time $\langle \tau\rangle$, standard deviation $\sigma$, KS distance for the fitted tail, and the burstiness ($B$) and memory ($M$) coefficients. For low-count flows and for sojourn-time rows, these values should be interpreted descriptively.\label{tab:summary}} \\

\toprule
Process & Flow & Data points & Exponent ($\gamma$) & $\tau_0$ (s) & $\langle\tau\rangle$ (s) & $\sigma$ (s) & KS & $B$ & $M$ \\
\midrule
\endhead
\midrule
\multicolumn{10}{r}{Continued on next page} \\
\midrule
\endfoot
\bottomrule
\endlastfoot

IBZ Arrivals & Global & 1403 & 2.19 & 1600.0 & 1821.3 & 7743.2 & --- & --- & --- \\
Shelf Rackings & Global & 26352 & 1.86 & 232.0 & 97.2 & 1461.1 & --- & --- & --- \\

OBZ Arrivals & Global & 13325 & 2.89 & 180.0 & 192.7 & 2354.4 & 0.05 & 0.85 & 0.07 \\
OBZ Arrivals & FactoryPickup$\to$CustomerPickup & 11 & --- & --- & 197225.5 & 227758.0 & --- & 0.07 & -0.00 \\
OBZ Arrivals & FactoryPickup$\to$DirectShip & 400 & 1.80 & 840.0 & 5661.6 & 45618.9 & 0.08 & 0.78 & 0.00 \\
OBZ Arrivals & FactoryPickup$\to$OutboundDrag & 1260 & 1.80 & 120.0 & 1852.0 & 17373.8 & 0.06 & 0.81 & 0.16 \\
OBZ Arrivals & FullLoadPrep$\to$CustomerPickup & 1042 & 2.14 & 14640.0 & 2170.5 & 27710.2 & 0.07 & 0.85 & -0.00 \\
OBZ Arrivals & FullLoadPrep$\to$DirectShip & 4470 & 2.12 & 4560.0 & 511.8 & 7379.1 & 0.06 & 0.87 & 0.39 \\
OBZ Arrivals & FullLoadPrep$\to$OutboundDrag & 2908 & 1.98 & 4680.0 & 799.2 & 10088.2 & 0.04 & 0.85 & 0.27 \\
OBZ Arrivals & InboundDrag$\to$CustomerPickup & 97 & --- & --- & 25858.1 & 45150.0 & --- & 0.27 & -0.06 \\
OBZ Arrivals & InboundDrag$\to$DirectShip & 383 & 2.75 & 9120.0 & 6529.5 & 18494.1 & 0.06 & 0.48 & 0.02 \\
OBZ Arrivals & InboundDrag$\to$OutboundDrag & 212 & --- & --- & 11816.9 & 28801.2 & --- & 0.42 & 0.02 \\
OBZ Arrivals & PickPrep$\to$CustomerPickup & 771 & 1.91 & 180.0 & 2916.9 & 27393.0 & 0.06 & 0.81 & -0.00 \\
OBZ Arrivals & PickPrep$\to$DirectShip & 3551 & 2.69 & 360.0 & 698.3 & 11202.3 & 0.03 & 0.88 & 0.02 \\
OBZ Arrivals & PickPrep$\to$OutboundDrag & 3743 & 2.72 & 360.0 & 621.8 & 10878.7 & 0.03 & 0.89 & 0.01 \\

OBZ Departures & Global & 3601 & 2.93 & 1796.0 & 698.3 & 5659.7 & 0.04 & 0.78 & 0.04 \\
OBZ Departures & FactoryPickup$\to$CustomerPickup & 11 & --- & --- & 178605.6 & 189161.7 & --- & 0.03 & -0.10 \\
OBZ Departures & FactoryPickup$\to$DirectShip & 396 & 2.20 & 9255.0 & 5160.1 & 28089.5 & 0.06 & 0.69 & 0.13 \\
OBZ Departures & FactoryPickup$\to$OutboundDrag & 405 & 2.24 & 2460.0 & 5705.6 & 25539.6 & 0.03 & 0.63 & 0.03 \\
OBZ Departures & FullLoadPrep$\to$CustomerPickup & 192 & --- & --- & 11892.7 & 45340.9 & --- & 0.58 & 0.15 \\
OBZ Departures & FullLoadPrep$\to$DirectShip & 520 & 2.26 & 3109.0 & 4426.2 & 27551.1 & 0.04 & 0.72 & 0.05 \\
OBZ Departures & FullLoadPrep$\to$OutboundDrag & 358 & 2.18 & 2640.0 & 6509.2 & 22064.2 & 0.05 & 0.54 & 0.27 \\
OBZ Departures & InboundDrag$\to$CustomerPickup & 129 & --- & --- & 19442.3 & 41021.5 & --- & 0.36 & -0.00 \\
OBZ Departures & InboundDrag$\to$DirectShip & 1710 & 1.73 & 659.0 & 1458.4 & 9102.8 & 0.05 & 0.72 & 0.05 \\
OBZ Departures & InboundDrag$\to$OutboundDrag & 332 & 2.05 & 2700.0 & 7568.1 & 22396.0 & 0.06 & 0.49 & 0.04 \\
OBZ Departures & PickPrep$\to$CustomerPickup & 327 & 1.76 & 1836.0 & 6999.5 & 34210.0 & 0.08 & 0.66 & 0.10 \\
OBZ Departures & PickPrep$\to$DirectShip & 919 & 2.23 & 2113.0 & 2691.2 & 22155.2 & 0.04 & 0.78 & 0.05 \\
OBZ Departures & PickPrep$\to$OutboundDrag & 387 & 2.24 & 2640.0 & 6021.4 & 26612.3 & 0.05 & 0.63 & 0.04 \\

OBZ Sojourns & Global & 37142 & 5.06 & 258960.0 & 83360.5 & 98355.8 & 0.04 & 0.08 & 0.49 \\
OBZ Sojourns & FactoryPickup$\to$CustomerPickup & 27 & --- & --- & 145746.4 & 90610.5 & --- & -0.23 & 0.47 \\
OBZ Sojourns & FactoryPickup$\to$DirectShip & 1669 & 5.63 & 381491.0 & 182659.2 & 153159.8 & 0.14 & -0.09 & 0.90 \\
OBZ Sojourns & FactoryPickup$\to$OutboundDrag & 4918 & 2.28 & 21360.0 & 44267.6 & 60352.9 & 0.04 & 0.15 & 0.55 \\
OBZ Sojourns & FullLoadPrep$\to$CustomerPickup & 1190 & 1.83 & 22192.0 & 88904.1 & 101412.8 & 0.11 & 0.07 & -0.01 \\
OBZ Sojourns & FullLoadPrep$\to$DirectShip & 5645 & 2.24 & 29495.0 & 62328.3 & 75477.3 & 0.07 & 0.10 & -0.02 \\
OBZ Sojourns & FullLoadPrep$\to$OutboundDrag & 3410 & 2.39 & 48420.0 & 59598.8 & 66095.2 & 0.08 & 0.05 & 0.01 \\
OBZ Sojourns & InboundDrag$\to$CustomerPickup & 271 & --- & --- & 188216.0 & 241876.9 & --- & 0.12 & 0.75 \\
OBZ Sojourns & InboundDrag$\to$DirectShip & 7477 & 5.74 & 299539.0 & 133111.9 & 116561.5 & 0.06 & -0.07 & 0.68 \\
OBZ Sojourns & InboundDrag$\to$OutboundDrag & 3001 & 9.15 & 259140.0 & 56528.5 & 68510.5 & 0.07 & 0.10 & 0.71 \\
OBZ Sojourns & PickPrep$\to$CustomerPickup & 910 & 2.72 & 105173.0 & 119984.4 & 112495.2 & 0.10 & -0.03 & -0.05 \\
OBZ Sojourns & PickPrep$\to$DirectShip & 4153 & 2.32 & 34070.0 & 74775.0 & 73234.8 & 0.09 & -0.01 & 0.01 \\
OBZ Sojourns & PickPrep$\to$OutboundDrag & 4471 & 2.23 & 33900.0 & 61093.8 & 69315.2 & 0.06 & 0.06 & -0.00 \\
\end{longtable}

\clearpage

\section{Scaling regimes summary}
\label{app:scaling}

To make the visual separation in Fig.~\ref{fig:scaling} explicit, we classify pallets as off-baseline when $\log(\tau_i/\tau_i^{\rm null})>\log 2$ and $\tau_i>24\,\mathrm{h}$, where $\tau_i^{\rm null}=A_i/\lambda$.

\begin{figure}[H]
\centering
    \includegraphics[width=1.0\textwidth]{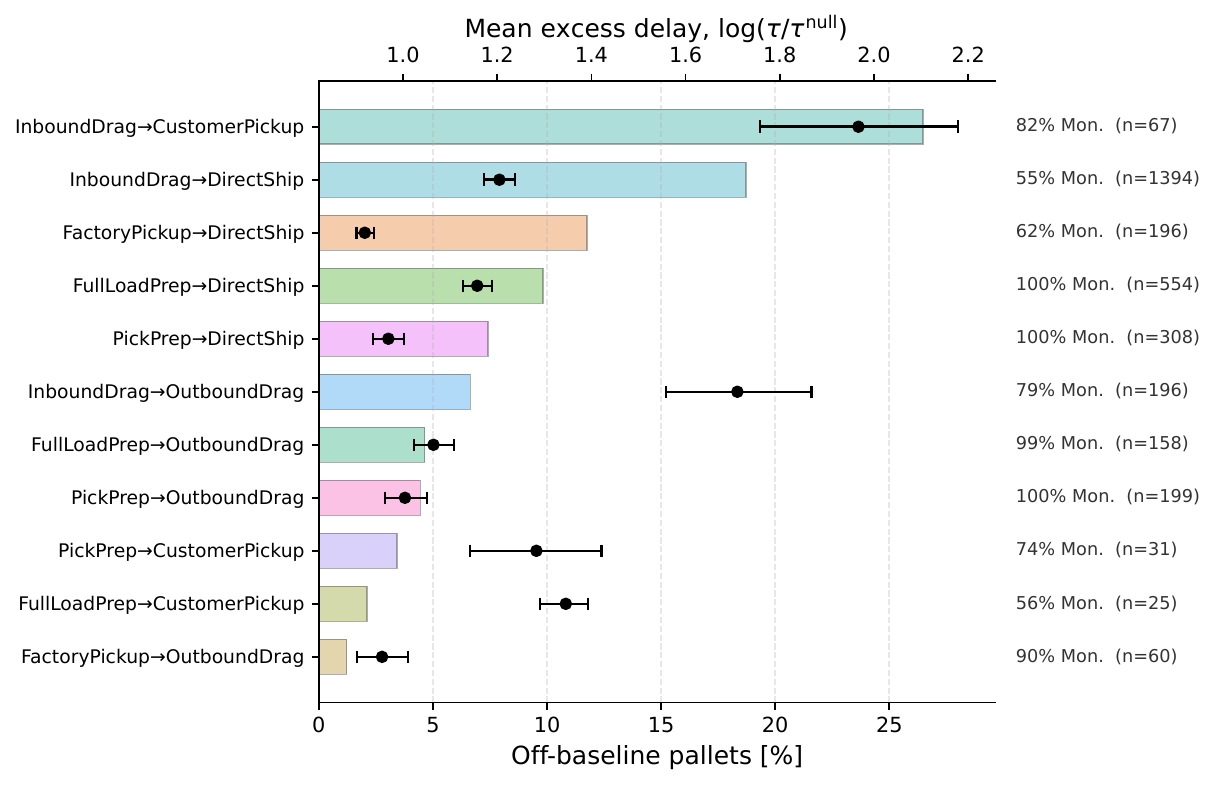}
    \caption{\textbf{Off-baseline activity-scaling statistics by logistics flow.} Bars show the fraction of pallets classified as off-baseline in each logistics flow. A pallet is classified as off-baseline when $\log(\tau_i/\tau_i^{\rm null})>\log 2$ and $\tau_i>24\,\mathrm{h}$, where $\tau_i^{\rm null}=A_i/\lambda$ is the activity-clock null expectation used in Fig.~\ref{fig:scaling}. Black points and horizontal intervals show the bootstrap mean and 95\% confidence interval of $\log(\tau_i/\tau_i^{\rm null})$ among off-baseline pallets. Right-hand labels report the share of off-baseline pallets departing on Monday and, in parentheses, the number of off-baseline pallets. The concentration of high Monday shares across several flows supports the interpretation that the off-baseline branch reflects scheduled weekend accumulation and Monday clearance rather than high concurrent activity alone.}
    \label{fig:summary}
\end{figure}

Figure~\ref{fig:summary} and Table~\ref{tab:scaling_summary} summarize the resulting residual branch by logistics flow. The off-baseline pallets are not uniformly distributed across routes. The largest fractions occur for \texttt{InboundDrag}$\to$\texttt{CustomerPickup} and \texttt{InboundDrag}$\to$\texttt{DirectShip}, indicating that inbound-transfer-origin flows are especially exposed to calendar-driven residence-time inflation. Among the off-baseline pallets, the mean residual is
positive in every reported flow, and the Monday share is high for most routes, often approaching or reaching 100\%, showing that the residual branch identified from the scaling relation is strongly associated with Monday clearance. Since Monday departure is not used in the classification rule, this provides an independent calendar interpretation of the two-regime structure observed in Fig.~\ref{fig:scaling}. Rows with very small off-baseline counts should be read only as qualitative support.

\begin{table}[H]
\centering
\small
\caption{\textbf{Numerical summary of off-baseline scaling regimes by logistics flow.}
For each flow, the table reports the total number of pallets, the number and
fraction classified as off-baseline, the Monday share among off-baseline pallets,
and the bootstrap mean and 95\% confidence interval of the residual
$\log(\tau_i/\tau_i^{\rm null})$ among off-baseline pallets.}
\label{tab:scaling_summary}
\begin{tabular}{lrrrrl}
\toprule
Flow & $n$ & $n_{\rm off}$ & Off-baseline (\%) & Monday off-baseline (\%) & $\langle \log(\tau_i/\tau_i^{\rm null})\rangle_{\rm off}$ \\
\midrule
InboundDrag$\to$CustomerPickup & 253 & 67 & 26.5 & 82.1 & 1.97 [1.76, 2.18] \\
InboundDrag$\to$DirectShip & 7447 & 1394 & 18.7 & 55.2 & 1.20 [1.17, 1.24] \\
FactoryPickup$\to$DirectShip & 1669 & 196 & 11.7 & 61.7 & 0.92 [0.90, 0.94] \\
FactoryPickup$\to$CustomerPickup & 27 & 3 & 11.1 & 100.0 & --- \\
FullLoadPrep$\to$DirectShip & 5645 & 554 & 9.8 & 100.0 & 1.16 [1.13, 1.19] \\
PickPrep$\to$DirectShip & 4153 & 308 & 7.4 & 100.0 & 0.97 [0.94, 1.00] \\
InboundDrag$\to$OutboundDrag & 2951 & 196 & 6.6 & 78.6 & 1.71 [1.56, 1.87] \\
FullLoadPrep$\to$OutboundDrag & 3410 & 158 & 4.6 & 98.7 & 1.06 [1.02, 1.11] \\
PickPrep$\to$OutboundDrag & 4471 & 199 & 4.5 & 100.0 & 1.00 [0.96, 1.05] \\
PickPrep$\to$CustomerPickup & 905 & 31 & 3.4 & 74.2 & 1.28 [1.14, 1.42] \\
FullLoadPrep$\to$CustomerPickup & 1190 & 25 & 2.1 & 56.0 & 1.35 [1.29, 1.39] \\
FactoryPickup$\to$OutboundDrag & 4915 & 60 & 1.2 & 90.0 & 0.96 [0.90, 1.01] \\
\bottomrule
\end{tabular}
\end{table}

\end{document}